%% file: ECOSTA.tex
\documentclass{article}

\usepackage[margin=1in]{geometry}
\usepackage[utf8]{inputenc}
\usepackage{authblk}
\usepackage{setspace}
\usepackage[linguistics]{forest}
\usepackage{color,soul}
\usepackage[colorlinks=TRUE,linkcolor=blue,citecolor=blue]{hyperref}
\usepackage{float}
\usepackage{amsmath,amsfonts,amsthm} 
\usepackage{graphicx,psfrag,epsf}
\usepackage{graphics}
\usepackage{enumerate}
\usepackage{mathtools}
\usepackage{tikz-qtree}
\usepackage{natbib}
\usepackage{pdfpages}
\usepackage{empheq}
\usepackage{placeins}
\usepackage{lscape}
\usepackage{forest}
\usepackage{textcomp}
\usepackage[normalem]{ulem}
\newtheorem{assumption}{Assumption}

\newcommand*{\bW}{{\mathbf{W}}}

\newcommand*{\bX}{{\mathbf{X}}}

\newcommand*{\bS}{{\mathbf{S}}}
\newcommand*{\bY}{{\mathbf{Y}}}
\newcommand*{\bG}{{\mathbf{G}}}

\newcommand*{\uS}{{\underline{S}}}
\newcommand*{\uY}{{\underline{Y}}}
\title{Bayesian principal stratification with longitudinal data and
truncation by death}
\author[1]{Giulio Grossi}
\author[2]{Marco Mariani}
\author[1]{Alessandra Mattei}
\author[3]{Fabrizia Mealli}
\affil[1]{Department of Statistics, Computer Science, Applications, University of Florence}
\affil[2]{IRPET - Istituto Regionale Programmazione Economica Toscana}
\affil[3]{European University Institute}
\date{}

\begin{document}
\maketitle
\setstretch{1.3}
\begin{abstract}
In many causal studies, outcomes are 'censored by death,' in the sense that they are neither observed nor defined for units who die. In such studies, the focus is usually on the stratum of 'always survivors' up to a single fixed time s. Building on a recent strand of the literature, we propose an extended framework for the analysis of longitudinal studies, where units can die at different time points, and the main endpoints are observed and well-defined only up to the death time. We develop a Bayesian longitudinal principal stratification framework, where units are cross-classified according to the longitudinal death status. Under this framework, the focus is on causal effects for the principal strata of units that would be alive up to a time point s irrespective of their treatment assignment, where these strata may vary as a function of s. We can get precious insights into the effects of treatment by inspecting the distribution of baseline characteristics within each longitudinal principal stratum, and by investigating the time trend of both principal stratum membership and survivor-average causal effects. We illustrate our approach for the analysis of a longitudinal observational study aimed to assess, under the assumption of strong ignorability of treatment assignment, the causal effects of a policy promoting start-ups on firms’ survival and hiring policy, where firms' hiring status is censored by death.
\end{abstract}

\section{Introduction}  
Start-up businesses can contribute significantly to urban regeneration and regional development through the creation of innovative and creative environments, see e.g: \cite{roman2013start}. In fact, existing literature shows that start-ups are a key driver of productivity growth, and economic renewal(\citealp{haltiwanger2013creates}, \citealp{dumont2016contribution}). Moreover, start-ups are also a viable solution to tackle unemployment. 

In recent years, policymakers have turned their attention to promoting self-employment by directly supporting the creation of new businesses. These policies aim to create and maintain a favorable environment for small businesses, including those that have been successful in the past but may be looking to reinvent themselves. The main objective of policies targeting promising entrepreneurial endeavors is to decrease unemployment by providing self-employment opportunities for aspiring business owners and potentially, additional hiring opportunities for the unemployed. However, it is not uncommon for start-ups to face challenges and fail during their early stages, e.g., due to economic conditions and the inexperience of the founders.

Building on a recent strand of the literature (\cite{bia2020}), we propose an extended framework for the analysis of longitudinal studies, where units can be censored at different time points, and the main endpoints are observed and well-defined only up to the censoring time. The motivating application is the evaluation of a program implemented by the Tuscan Region to foster entrepreneurship through public guarantee for bank loans. The outcome of interest is the probability of observing hirings in the post-treatment periods. Evaluation of public programs can eventually be harmed by post-treatment complications. Outcomes can be censored by death, in the sense that they are neither observed nor defined for units who die.

This truncation can harm the comparison between treated and control arms' outcomes, as the truncation mechanism is non-ignorable (\cite{lin1996comparing}). The first solution for these complications spans from the work by \cite{heckman1976common} and \cite{heckman1979sample}, who see this problem as a sample selection situation and provide a structural equation modeling approach to deal with it. Early research focuses on comparing outcomes for surviving units under treatment and control, conditioning on covariates (\cite{decker1995international}), imputing the outcome for truncated units in the counterfactual scenario in which there hasn't been truncation (\cite{mcmahon2001joint}), and modeling the censoring mechanism (\cite{wu1988estimation} and \cite{fairclough1998missing})  

In our study, we address the issue of truncation due to death through the principal stratification approach. By leveraging structural assumptions and a Bayesian model-based method for inference, we tackle identification challenges. This approach, initially proposed by \cite{frangakis2002}, see also \cite{zhang2008evaluating}, \cite{mattei2007}, and \cite{mealli2012}, forms the basis of our analysis.
The Bayesian approach for causal inference, introduced by \cite{rubin1978} allows us to estimate flexibly the causal effects even in the presence of complex structures as mixtures of observed outcomes, with a clear interpretation in terms of causal quantities. This method has been formalized by \cite{imbens1997} for dealing with problems of noncompliance. It has been further extended into the principal stratification approach, often employed in the literature to treat with censoring or truncation by death problems, with non-ignorable missingness of the outcomes, see for instance \cite{zhang2003estimation}, \cite{chiba2011simple}, \cite{mattei2011}, \cite{mealli2012}, \cite{frumento2012evaluating}.

The contribution of this work is twofold: we aim to contribute to the existent literature about causal effect estimation in the presence of post-treatment complications by expanding the framework developed by \cite{bia2020}. We generalize to multiple periods the principal stratification approach for truncated outcomes and estimating causal quantities leveraging Bayesian inference. Moreover, we wish to contribute to the thematic literature about the evaluation of public support for start-ups, which has on several occasions focused on outcomes whose availability may depend on firm survival but has almost always done so by disallowing the problem, without recognizing and discussing its implications to draw causal claims. 

The work proceeds as follows: section \ref{sec:motivating} illustrates our motivating application, its main complication and the relevant literature on the applied field. Section \ref{sec:methodology} describes the methodological framework and the empirical strategy to estimate causal effects. Section \ref{sec:results} presents the main findings of the work. Section \ref{sec:conclusion} concludes. 

\FloatBarrier
\section{The puzzle of start-up subsidies}\label{sec:motivating}

Firm development during the first stages of activity can be harmed by the existence of credit or other finance-related constraints.
Financing challenges, particularly for new ventures, are significant due to limited relational capital and asymmetric information about the entrepreneurial project potential,  an uncertainty that may discourage possible lenders \citep{peneder2008problem, colombo2007funding, nigam2020impact}. In such a context, support programs can be vital for the survival of new firms. \\
The evaluation of start-up support programs is a hotly debated topic in the economic policy literature. On the one hand, if one believes the main goal of these programs is to guarantee the self-employment of their founders, there is consensus on the ability of these programs to prevent founders, through self-employment especially when they belong to 'vulnerable' strata such as youth and females, from a difficult stay in the regular labor market \cite[e.g.][]{kane2010importance, decker2014role, caliendo2014regional}.

However, from one takes entrepreneurship policy perspective, the judgment on such policies may be less positive, based on the fact that one is often faced with small business projects that lack relevant potential and, therefore, require allocating public support to enterprises of dubious efficiency \citep{shane2009encouraging, lukevs2019business}. 
 As shown in surveys of the international literature \citep{caliendo2016,dvoulety2016}, firms’ survival is usually the most important outcome considered in the existing evaluation studies, because it corresponds to the length of self-employment. This has also been the case in studies that, like ours, concern start-up programs implemented in Italy. For example, focusing on the same policy we analyze here, \cite{mariani2019} assesses the causal effect of public support on the hazard of exit of supported firms, finding that public support has indeed positive but unfortunately quite ephemeral effects. Concerning fairly similar programs \cite{mealli2001analisi} and \cite{battistin2001subsidised} also find positive, albeit short-lived, effects on survival, which reveal the low potential of the entrepreneurial projects supported. On the contrary, \citet{manaresi2021supporting} detect no effect on survival chances for new firms participating in the recent Italian "Start-up Act" national program.\\
Acknowledging criticisms from entrepreneurial policy experts, some studies have focused not only on the length of self-employment but also on other outcomes that may have social significance. To this regard, it has been assessed whether start-up programs yield a ‘double dividend’ by ensuring further job creation or, more seldom, innovation, with still partly mixed results \cite[e.g.][]{caliendo2016, caliendo2015subsidized, mariani2019, manaresi2021supporting}.  
It is precisely the analyses that examine whether start-up programs have resulted in double dividends that potentially suffer from the truncation-by-death problem for outcomes related to these dividends. Depending on how they are defined, outcomes after the death of the firm may or may not be truncated. For example, the number of jobs created by a dead enterprise is zero and therefore does not suffer from truncation. In contrast, the decision to hire or not to hire staff cannot be made in an enterprise that no longer exists, and therefore, the corresponding outcome is truncated. In previous studies drawing causal claims on double dividends, very little is said about how the corresponding outcomes are shaped after the death of firms, leaving much room for the reader's imagination. The only study that explicitly recognizes the potential truncation problem is \cite{mariani2019}, addressed by introducing assumptions of sequential ignorability of firm deaths conditional on covariates and treatment and outcome histories. Nevertheless, the authors themselves eventually admit that an approach based on principal stratification could be preferably used since such an approach acknowledges the endogeneity of firm closures and avoids the introduction of cumbersome assumptions such as sequential ignorability. Illustrating what a principal stratification approach can add to causal analyses on the double dividend of start-up subsidies is the main purpose of our paper.
\FloatBarrier

	\subsection{The "Fare impresa" program in Tuscany, Italy}\label{sec:data}
	
Our work has been motivated by the evaluation of the program called ``Fare Impresa'' (Doing Business)  implemented in Tuscany, a region in Central Italy, between 2011 and 2015. 
The main objective of the program is to foster entrepreneurship of young and female businesses in the very first period of activity through bank loans assisted by public guarantees. 

Participation in the Doing Business program is subject to some access criteria: 
the firm's owner needs to be female of any age or male aged 18-40 and the firm must satisfy age constraints.
Newly established firms (no older than two years, or starting activity within six months) and established firms seeking expansion opportunities with less than 5 years of activity are eligible to participate in the program. 
Firms entering the program are eligible to receive a public-assisted guarantee that should help them obtain bank loans at a reduced interest rate to start or grow their businesses.
The loans can last up to ten years, with the guarantee covering up to 80\% of the requested amount.
A regional financial intermediary oversees the loan request process and ensures that it is conducted correctly. 
In total, 1837 firms receive the guarantee backed by public authorities, out of which 1563 (85.1\%) receive bank credit, and 274 (14.9\%) have credit rejections. 

Our dataset is constructed by merging three data sources. Firstly, we have the records of a regional financial intermediary (Fidi Toscana), including information on the start date for the firm's activity, the business sector, the location of the investment, and demographic characteristics of the firm's owner, such as age and gender. Secondly, we collect data on the firms' end dates from the Chamber of Commerce archives. Lastly, we gather hiring information from the Tuscan Job Information system, which also provides information on the type of contract and its duration.
The dataset we obtain by merging these data sources includes for each firm a wealth of information on several background characteristics, the bank's decision over the credit concession, and the hiring decision and the survival status for three years after the bank's decision over the credit concession, namely, for the years 2012, 2013 and  2014.
Formally, for each firm $i$,  $i=1, \dots, N$, $N=1837$,  we observe 
 a vector of time-invariant and firm-specific pre-treatment covariates $\bX_i$ including information on the variables shown in Table~\ref{tab:des}; and a binary treatment indicator, $W_i$, for the granting of a loan from a bank, with $W_i=1$ if firm $i$ receives a bank credit to promote their projects, and  $W_i=0$ otherwise.
 Moreover, in each of the three post-treatment periods $t\in \{1,2,3\}$, corresponding to years 2012, 2013 and  2014, for each firm $i$, we observe its survival status,$S_{i,t}$, and its
 hiring decision, $Y_{i,t}$.  The survival status,  $S_{i,t}$,  is a binary variable with $S_{i,t}=1$ if firm $i$ does not cease activity during year $t$ and zero otherwise.
If firm $i$ is still alive at time $t$ --  $S_{i,t}=1$ -- then  $Y_{i,t}$, is equal to one if firm $i$ forms employment contracts during year $t$, and zero otherwise.  
If firm $i$ is ceased at  time $t$ --  $S_{i,t}=0$ --then  its
hiring decision $Y_{i,t}$ is truncated by death, in the sense that it is neither observed nor defined, and we set it equal to $\ast$, a non-real value.  
Let $\bW$, $\bY_t$, $\bS_t$, $t \in (1,2,3) $, be $N$-dimensional vectors with $i$th entries equal to $W_i$, $Y_{i,t}$ and $S_{i,t}$, respectively, and let $\bX$ be a $N\times K$ matrix of pre-treatment variables, with the $i$-th row equal to $\bX_i$.

Table \ref{tab:des} shows descriptive statistics on the pre-treatment variables for the whole sample grouped by treatment. 	Around three forty of the firms in the study are owned by young individuals (76.5\%), while female owners make up 57.1\%. The majority of firms are new start-ups (92.1\%), and 60.6\% are owned by a single individual. Around half of the loans are granted by a local bank, and only 19.8\% of the projects are located in urban areas. This suggests that the program has been particularly focused on rural areas in Tuscan, where local banks are more common. Additionally, 25.1\% of the firms have already hired workers at the beginning of the observation period. 	The firms participating in the program represent a range of different industries, with 11.1\% of the sample involved in manufacturing, 32.7\% engaged in retail, 27.6\% in hospitality, and 12.2\% in service to person (beauty and hairdressing). The remaining portion of the sample represents other types of economic activities.

Our study is an observational causal study in the sense the mechanism underlying the bank's decision over the credit concession is unknown.
Confounding bias is a key challenge when estimating causal effects from observational data.
Indeed, we observe significant discrepancies in characteristics between treated and control firms.
The subsample of firms that receive a bank credit comprises, on average, a lower proportion of sole-ownership firms, a higher proportion of firms that 
submit the credit request to a local bank, a higher proportion of
hospitality and personal service industries, and a   lower proportion of firms working in sectors other than the manufacturing, retail, hospitality, and personal service sectors.
Therefore the bank's decision over the credit concession appears to be influenced by firm-specific factors, including the business ownership structure and the firm owner's relationship capital.  Local banks generally favor funding projects within their community, benefiting smaller businesses. In contrast, national banks adhere to more stringent lending criteria, leading to a higher rejection rate for applications that fail to meet these thresholds.

	\begin{table}
		\caption{Descriptive statistics for background covariates} 
		\centering
		\label{tab:des}
			\begin{tabular}[t]{lccc }
				\hline
			 & \multicolumn{3}{c}{Mean} \\
				\cline{2-4}
	Variable	& $W_i=0$ & $W_i=1$ & Overall \\
				\hline
				Young & 0.770 & 0.760 & 0.765 \\
				Female & 0.540 & 0.576 & 0.571\\
				Has employees & 0.226 & 0.257 & 0.251 \\
				Start-up & 0.901 & 0.925 & 0.921\\
				Sole-ownership & 0.657 & 0.597 & 0.606 \\
				Sector of activity\\
			\quad 	Manufacturing         & 0.131 & 0.107 & 0.111 \\
				\quad 	Retail            & 0.296 & 0.332 & 0.327\\
				\quad 	Hospitality       & 0.226 & 0.285 & 0.276 \\
				\quad 	Service to person & 0.080 & 0.130 & 0.122 \\
				\quad  Other              & 0.273 & 0.146 & 0.164 \\
				Local bank & 0.401 & 0.553 & 0.531 \\
				Urban location & 0.197 & 0.198 & 0.198 \\
				\hline
			\end{tabular}
	\end{table}
Table~\ref{tab:hire} shows descriptive statistics of the post-treatment variables, firm's survival status and hiring decision in each of the three post-treatment times.
In our sample, 185  firms (10.1\%) cease the activity by the end of the observation period.
The proportion of firms that cease the activity is higher among control firms than among treated firms (0.299 vs. 0.122 ), but the proportion of closing firms is decreasing over time among firms that do not receive the credit, while it is increasing for the treated ones. In each post-treatment time point, but first, the proportion of firms that form new contracts is higher in the subsample of the treated firms. 
These naive comparisons between treated and control firms are not directly interpretable in causal terms due to the divergences in covariate distributions:  we must carefully adjust treatment comparison for the difference in the covariates.
 Moreover, we need to take into account that hiring decisions is truncated by firms' deaths in drawing inferences on the causal effects of it. 
Table \ref{tab:groups} shows the data structure and the proportion of firms in each subgroup defined by the observed survival status over time and treatment status.
We find that 95.54\% of the treated firms are still active at the end of the observation period. 
In the control group, the composition is more varied with only 47.76\% of firms still active at the end of the observation period and more than 13\%, 18\% and 20\% of firms that close in the first, second and third observation year, respectively.

	\begin{table}[h]\label{tab:hire}
		\caption{Descriptive statistics of the post-treatment variables, firm's survival status and hiring decision, for each of the three post-treatment times}
		\centering
		
		\begin{tabular}[t]{l|cc|cc|cc}
			\hline
		  &\multicolumn{2}{|c|}{\textbf{t=1}}  & \multicolumn{2}{c|}{\textbf{t=2}} & \multicolumn{2}{c}{\textbf{t=3}}\\
			\multicolumn{1}{c|}{\textbf{Variable}}	& $W_i=0$ & $W_i=1$ & $W_i=0$ & $W_i=1$ & $W_i=0$ & $W_i=1$ \\
			\hline
 Survival Status ($S_{i,t}$) &&&&&&	\\
	\qquad	  Active & 0.807 & 0.978 & 0.730 & 0.929 & 0.701 & 0.877\\
			\qquad		Closed & 0.193 & 0.022 & 0.077 & 0.049 & 0.029 & 0.052\\
	&&&&&&\\
	Hiring decision ($Y_{i,t}$)&&&&&&	\\	
\qquad 	Hiring & 0.628 & 0.598 & 0.591 & 0.655 & 0.599 & 0.707\\
	\qquad  No-Hiring & 0.179 & 0.380 & 0.139 & 0.274 & 0.102 & 0.171\\
\qquad		$\ast$ & 0.193 & 0.022 & 0.270 & 0.071 & 0.299 & 0.122\\
			\hline
		\end{tabular}
	\end{table}

\begin{table}[ht]
\centering
\caption{Data structure and observed proportion of groups defined by the survival status over time and the treatment status ($\checkmark$ means that the outcome is definite, $\ast$ stands for truncated outcome).}
\begin{tabular}{c|ccc|ccc|cc}
\hline
\multicolumn{1}{l|}{W} & \multicolumn{1}{l}{$S_{i,1}$} & \multicolumn{1}{l}{$S_{i,2}$} & \multicolumn{1}{l|}{$S_{i,3}$} & \multicolumn{1}{l}{$Y_{i,1}$} & \multicolumn{1}{l}{$Y_{i,2}$} & \multicolumn{1}{l|}{$Y_{i,3}$} &   \multicolumn{1}{l}{Group Proportion} \\
\hline
1     & 0     & 0     & 0   & $\ast$     &$ \ast  $   & $\ast$    & 2.44\% \\
1     & 1     & 0     & 0     &   $\checkmark$   & $\ast$     & $\ast$  & 1.46\% \\
1     & 1     & 1     & 0     & $\checkmark$     & $\checkmark$     & $\ast$ & 0.56\% \\
1     & 1     & 1     & 1     & $\checkmark$     & $\checkmark$     & $\checkmark$ & 95.54\% \\
\hline
0     & 0     & 0     & 0    & $\ast$     & $\ast$     & $\ast$  & 13.18\% \\
0     & 1     & 0     & 0     & $\checkmark$     & $\ast$     & $\ast$  & 18.91\% \\
0     & 1     & 1     & 0    & $\checkmark$     & $\checkmark$     & $\ast$ & 20.15\% \\
0     & 1     & 1     & 1    & $\checkmark$     & $\checkmark$     & $\checkmark$  & 47.76\% \\
\hline
\end{tabular}%
\label{tab:groups}%
\end{table}%

	\FloatBarrier

\section {Potential outcomes and causal estimands}\label{sec:methodology}
	
The causal estimands are defined using the potential outcome approach \citep{rubin1974}, under the Stable Unit Treatment Value Assumption \cite[SUTVA,][]{rubin1980}, which rules out hidden versions of the treatment and interference between units.
SUTVA appears a reasonable assumption in our study. We can consider the treatment as homogeneous across the different firms, as all the businesses are similar in size, and therefore, we could expect similar entrepreneurial projects with similar financed amounts. 
We can also safely assume no interference between units, given that treated businesses are scattered through Tuscany, and thus, we could expect low interaction between them. Moreover, it is unlikely that the granting of a loan to a firm may have a sizeable effect on the hiring decisions of other firms.  
	
Under SUTVA, for each firm $i$ and time $t$ we define the following pairs of potential outcomes for  firm $i$'s  survival status and  hiring decision: 
$S_{i,t}(1)$ and $S_{i,t}(0)$, which are the values of firm $i$'s survival status at time $t$ if the firm was assigned to treatment and to control, respectively; and $Y_{i,t}(1)$ and $Y_{i,t}(0)$, which are the values of firm $i$'s hiring decision at time $t$ if the firm was assigned to treatment and to control, respectively. 
Let  $\underline{S}_{i,t}=(S_{i,t}(0),S_{i,t}(1))$ and $\underline{Y}_{i,t}=(Y_{i,t}(0),Y_{i,t}(1))$ denote the row vectors	of potential outcomes for firm $i$'s survival status and hiring decision   at time $t$, and let $\underline{\bS}_{t}=(\bS_{t}(0),\bS_{t}(1))$
be $\underline{\bY}_{t}=(\bY_{t}(0),\bY_{t}(1))$  the $N \times 2 $ matrices of potential outcomes at time $t$ with $i$th row equal to $\underline{S}_{i,t}$ and $\underline{Y}_{i,t}$, respectively.

\FloatBarrier

\subsection{Principal stratification Approach}\label{sec:strata}
	
Assessing the causal effects of credit on the
hiring decision is problematic because hiring decision at time $t$ can only be observed for firms that are still active at time $t$ ($S_{i,t} = 1$); hiring decision at time $t$  is not only unobserved but also undefined when  $S_{i,t} = 0$, namely, it is censored by death. In principle, a causal estimand of interest would be the effect of the treatment on the hiring decision for those firms who would be active under both treatment statutes ($Y_{i,t}(0)=Y_{i,t}(1)=1$).
 
We deal with this issue using the principal stratification approach, firstly proposed by \cite{frangakis2002}, see also \cite{rubin2006}, \cite{forastiere2016identification}, \cite{mattei2007}, \cite{bia2020} and \cite{frumento2012}. 
Principal stratification allows us to classify units into latent groups, the principal strata, defined by the joint potential outcomes of the intermediate variable under each of the treatments, $(S_{i,t}(0), S_{i,t}(1))$.
	
In our study, in principle, at each time point $t$, $t=1,2,3$, we can classify firms into four latent groups with respect to their survival status under treatment and control:
	\textit{Never Survivors} -- $NS=\{i: S_{i,t}(0)=S_{i,t}(1)=0\}$ -- businesses that would cease their activity by time $t$, irrespective of their treatment assignment;    \textit{Always Survivors} -- $AS=\{i: S_{i,t}(0)=S_{i,t}(1)=1\}$ --  businesses that would continue their activity irrespective of their treatment assignment; 
\textit{Compliant Survivors} -- $CS=\{i: S_{i,t}(0)=0, S_{i,t}(1)=1\}$ --   businesses that would continue their activity, \textit{if} they were assigned to the treatment but would cease their activity if assigned to control; and  \textit{Defiant Survivors} -- $DS=\{i: S_{i,t}(0)=1, S_{i,t}(1)=0\}$ --   businesses that would cease their activity \textit{if} they received the treatment but would continue their activity if were assigned to control. 
Let $G_{i,t}$ be the indicator for the principle stratum membership of firm $i$ at time $t$, $G_{i,t} \in \{NS, DS, CS, AS\}$.
	
Leveraging the longitudinal structure of our data, 	we can classify firms with respect to the joint value of the potential survival status in each post-treatment period. Formally, the longitudinal principal stratum membership for each firm $i$ is defined as follows \citep{bia2020}: 
$$(S_{i,1}(0), S_{i,1}(1), S_{i,2}(0), S_{i,2}(1), S_{i,3}(0), S_{i,3}(1)).$$
Let	$ \bG_i=(G_{i,1}.G_{i,2}.G_{i,3})$ be the indicator for the longitudinal principal stratum membership.
Because death is an absorbing event, that is, $S_{i,t}(W_i) = 0$ implies that $S_{i,t'}(W_i)= 0$ for all $t' \geq t$, some longitudinal principal strata do not exist by construction. For instance, a firm that belongs to the principal stratum of the compliant survivors at time $t$, $G_{i,t}=CS$, can only either stay in the stratum of compliant survivors or transit to the stratum of never survivors at time $t+1$; $G_{i,t+1} \in \{CS, NS\}$. 
Therefore, firms can be classified into  16 longitudinal principal strata as shown in Table \ref{tab:strata1}. 
\begin{table}[t]
\centering
\caption{Longitudinal principal strata}
\label{tab:strata1}%
$$
\begin{array}{cc cc cc c}
\hline
S_{i,1}(0)& S_{i,1}(1)& S_{i,2}(0)& S_{i,2}(1)& S_{i,3}(0)& S_{i,3}(1) &\bG_i \\
\hline
1 & 1 & 1 & 1 & 1&1 & AS.AS.AS\\
1 & 1 & 1 & 1 & 0&1 & AS.AS.CS\\
1 & 1 & 1 & 1 & 1&0 & AS.AS.DS\\
1 & 1 & 1 & 1 & 0&0 & AS.AS.NS\\
1 & 1 & 0 & 1 & 0&1 & AS.CS.CS\\
1 & 1 & 0 & 1 & 0&0 & AS.CS.NS\\
1 & 1 & 0 & 0 & 0&0 & AS.NS.NS\\
1 & 1 & 1 & 0 & 1&0 & AS.DS.DS\\
1 & 1 & 1 & 0 & 0&0 & AS.DS.NS\\ 
0 & 1 & 0 & 1 & 0&1 & CS.CS.CS\\
0 & 1 & 0 & 1 & 0&0 & CS.CS.NS\\
0 & 1 & 0 & 0 & 0&0 & CS.NS.NS\\ 
1 & 0 & 1 & 0 & 1&0 &  DS.DS.DS\\
1 & 0 & 1 & 0 & 0&0 &  DS.DS.NS\\
1 & 0 & 0 & 0 & 0&0 &   DS.NS.NS\\
0 & 0 & 0 & 0 & 0&0 &  NS.NS.NS\\
\hline	
\end{array}
$$
\end{table}%
	
	\subsection {Survivor average causal effects}\label{sec:cate}
The principal stratification framework makes it clear that some potential outcomes for hiring decisions are not defined for some principal strata: in each time point $t$, 
 $Y_{i,t}(0)=\ast $ for compliant survivors and never survivors; and  
 $Y_{i,t}(1)=\ast $ for defiant survivors and never survivors.  
In each time point $t$, only for the subgroup of always survivors, both potential outcomes for hiring decision, $Y_{i,t}(0)$ and $Y_{i,t}(1)$ are defined. Therefore, a well-defined real value for the average causal effect of treatment versus control on hiring decisions exists only for the group of always survivors. These types of effects are called
survivor average causal effects, SACEs \citep{rubin2006}.
In our longitudinal setting, where $G_{i,t}=AS$ implies that $G_{i,t'}=AS$ for all $t' \leq t$, we can defined   the following survivor average causal effects:
\begin{eqnarray} 
 	SACE_{1}(1)&=& \mathbb{E}[(Y_{i,1}(1)-Y_{i,1}(0)\mid G_{i,1}=AS)] 
 	\label{eq:SACE1}\\
 			SACE_{1:2}(t)&=& \mathbb{E}[(Y_{i,t}(1)-Y_{i,t}(0)\mid G_{i,1}=AS, G_{i,2}=AS)] \quad t=1,2 \label{eq:SACE2}\\
 			SACE_{1:3}(t) &=& \mathbb{E}[(Y_{i,t}(1)-Y_{i,t}(0) \mid G_{i,1}=AS, G_{i,2}=AS, G_{i,3}=AS)]\quad t=1,2,3 \label{eq:SACE3}
 		\end{eqnarray}
$SACE_{1:t}(t')$, $t,t' \in {1,2,3}$ with $t' \leq t$ is the average causal effect at time $t'$, for the subpopulation of firms that would survive under both treatment and control at least up to time $t$.
 
These estimands may provide useful insights on the evolution of the causal effects over time.  In particular, comparing  $SACE_{1:t}(t')$ through time $t'$ can help us understand the short and mid-term effects of the policy. 
	
\subsection{Observed and missing potential outcomes} 
The fundamental problem of causal inference implies that we can only observe the outcome for both the intermediate variable (survival status) and the main endpoint (hiring decision) under the treatment received, while the other potential outcomes are missing. We observe
$$
S_{i,t}= W_i S_{i,t}(1) +(1-W_{i})S_{i,t}(0)\qquad and \qquad	Y_{i,t}= W_i Y_{i,t}(1) +(1-W_{i})Y_{i,t}(0),
$$	
but 
$$
S^{mis}_{i,t}= (1-W_i) S_{i,t}(1) + W_{i} S_{i,t}(0)\qquad and \qquad	Y^{mis}_{i,t}= (1-W_i) Y_{i,t}(1) +W_{i}Y_{i,t}(0).
$$
are missing.
Therefore, we do not generally observe the principal stratum membership for any unit, but each observed group defined by the treatment status and the survival status is a mixture of four longitudinal principal strata as shown in the upper panel in Table~\ref{tab:obs_strata}.
\begin{table}[t]
	\centering
	\caption{Observed data and corresponding longitudinal principal strata }
	\label{tab:obs_strata}%
 	$$
	\begin{array}{c ccc cccc}
		\hline
		W_{i,t}& S_{i,1} & S_{i,2} &  S_{i,3} &\multicolumn{4}{c}{\bG_i= G_{i,1}.G_{i,2}.G_{i.3} \hbox{ (No monotonicity)}} \\
		\hline
		1 & 1 & 1 & 1  & AS.AS.AS & AS.AS.CS & AS.CS.CS & CS.CS.CS \\
		1 & 1 & 1 & 0  & AS.AS.DS & AS.AS.NS & AS.CS.NS & CS.CS.NS\\
		1 & 1 & 0 & 0  & AS.NS.NS & AS.DS.DS & AS.DS.NS & CS.NS.NS\\
		1 & 0 & 0 & 0  & DS.DS.DS &  DS.DS.NS& DS.NS.NS &  NS.NS.NS\\
		\\
		0 & 1 & 1 & 1  & AS.AS.AS & AS.AS.DS & AS.DS.DS & DS.DS.DS \\
		0 & 1 & 1 & 0  & AS.AS.CS & AS.AS.NS & AS.DS.NS & DS.DS.NS \\
		0 & 1 & 0 & 0  & AS.NS.NS & AS.CS.CS & AS.CS.NS & DS.NS.NS\\
		0 & 0 & 0 & 0  & CS.CS.CS &  CS.CS.NS& CS.NS.NS &  NS.NS.NS\\
		\hline	
	\end{array}
	$$
		$$
	\begin{array}{c ccc cccc}
		\hline
		W_{i,t}& S_{i,1} & S_{i,2} &  S_{i,3} &\multicolumn{4}{c}{\bG_i= G_{i,1}.G_{i,2}.G_{i.3} \hbox{ (Under monotonicity)}} \\
		\hline
		1 & 1 & 1 & 1  & AS.AS.AS & AS.AS.CS & AS.CS.CS & CS.CS.CS \\
		1 & 1 & 1 & 0  & AS.AS.NS & AS.CS.NS & CS.CS.NS\\
		1 & 1 & 0 & 0  & AS.NS.NS  & CS.NS.NS\\
		1 & 0 & 0 & 0  &  NS.NS.NS\\
		\\
		0 & 1 & 1 & 1  & AS.AS.AS   \\
		0 & 1 & 1 & 0  & AS.AS.CS & AS.AS.NS   \\
		0 & 1 & 0 & 0  & AS.NS.NS & AS.CS.CS & AS.CS.NS  \\
		0 & 0 & 0 & 0  & CS.CS.CS &  CS.CS.NS& CS.NS.NS &  NS.NS.NS\\
		\hline	
	\end{array}
	$$
\end{table}%

Drawing inferences on the survivor causal effects of interest requires disentangling these mixtures, which is a particularly challenging task.
We deal with the inferential challenges using a Bayesian approach under some structural assumptions.

\section{Structural assumptions}\label{sec:assumptions}
We first posit a treatment assignment mechanism,  the process that determines which firms receive which treatments, and so which potential
 outcomes are realized and which are missing.
 In our application, firms voluntarily participate in a public policy program, and thus, the treatment assignment mechanism is also unknown and not under our control. We invoke the strong ignorability assumption \citep{rosenbaum1983}:
%
	\begin{assumption}{\textbf{Strong Ignorability of treatment assignment}}\label{ass:strong}
		\begin{itemize}
			\item Unconfoundness: $Pr(W_i\mid \uS_{i,1}, \uS_{i,2},\uS_{i,3}, \uY_{i,1}, \uY_{i,2}, \uY_{i,3}, \bX_i)=Pr(W_i \mid \bX_i)$
			\item Overlap: $0<Pr(W_i=1 \mid \bX_i)<1$
		\end{itemize} 
	\end{assumption}
Strong ignorability consists of two parts: Unconfoundness and overlap. The unconfoundness assumption rules out the presence of unmeasured confounders implying that within subpopulations of units with the same value for the covariates the treatment has been randomly assigned.
 This assumption is not testable, but we judge it reasonable in our study because we observe the same information that the bank uses to grant credit to firms. 
Overlap means that, within cells defined by the covariates, there are both treated and control units, at least in large samples. 
Strong ignorability implies the conditional distribution of the principal stratum membership given the covariates is the same in both treatment arms.
	
We assume monotonicity, that there are no defiant survivors, firms that would cease the activity if received the bank loan but would survive if they did not receive it:
\begin{assumption}{\textbf{Monotonicity}}
\label{ass:monotonicity} For all $i$ 
$$S_{i,t}(1)\geq S_{i,t}(0)  \qquad t=1,2,3 $$
\end{assumption}
Monotonicity appears very plausible in our application, where defiant survivors would be firms that would cease the activity if they received the bank loan, but would survive if they did not receive the bank loan. It seems implausible that such a group of firms exists; even because a business that receives a bank loan starts to refund it in the seventh year. 
	
Assumption \ref{ass:monotonicity}  implies that the number of possible principal strata reduces to ten, simplifying the observed mixture data structure as shown in the bottom panel in Table \ref{tab:obs_strata}.
	
Even under Assumptions \ref{ass:strong} and \ref{ass:monotonicity}, neither the longitudinal principal stratum proportions nor the survivor average causal effects in Equations \eqref{eq:SACE1}-\eqref{eq:SACE3} are fully non-parametrically identifiable.
We propose to deal with identification issues by using a flexible Bayesian parametric approach, which is often adopted in principal stratification analysis where inference involves techniques for incomplete data  \citep[e.g.,][]{ricciardi2020bayesian,bia2020}.

	\section{Bayesian Inference}\label{sec:bayesian}

	The model-based Bayesian approach, introduced by \cite{rubin1978}, is a comprehensive and versatile framework for analyzing complex data with missing information.
	Conceptually, the Bayesian approach does not require full identification:
	Bayesian inference is based on the posterior distribution of the parameters of interest, which is always proper if the prior distribution is proper.   The posterior distribution is derived by updating the prior distribution via the likelihood, irrespective of whether the parameters are fully or partially identified \cite[e.g.,][]{li2023bayesian}. Nevertheless, in a finite sample, posterior distributions of partially identified parameters may be weakly identifiable in the sense that they may have substantial regions of flatness 
	\cite[e.g.,][]{imbens1997, gustafson2010bayesian, ricciardi2020bayesian}.
Recent results also show that the joint distribution of the potential intermediate variables (the survival status) can be identified by using an outcome model for the potential outcomes for the primary endpoint (hiring decision) \citep{antonelli2023principal}.

	The general structure for conducting Bayesian causal inference with principal stratification was first outlined by \cite{imbens1997}. It was initially developed to address issues of all-or-none treatment noncompliance but has since been extended and applied in various contexts.  Our work builds on previous research in this area, including contributions by  \cite{mattei2007}, \cite{ricciardi2020bayesian} and \cite{bia2020}.

Fifteen quantities are associated with each firm $i$, $\bX_i$, $W_i$, and 
$S_{i,t}(0)$, $S_{i,t}(1)$, $Y_{i,t}(0)$, $Y_{i,t}(1)$ for $t=1,2,3$, where $\bX_i$, $W_i$, and  $S_{i,t}(W_i)$ and $Y_{i,t}(W_i)$, $t=1,2,3$,
are observed but $S_{i,t}(1-W_i)$ and $Y_{i,t}(1-W_i)$, $t=1,2,3$, are missing.	
Bayesian inference views all these quantities as random variables and centers around specifying a joint model for them:
$$P(\underline{\bS}_{1},\underline{\bS}_{2},\underline{\bS}_{3},\underline{\bY}_{1}, \underline{\bY}_{2}, \underline{\bY}_{3}, \textbf{W}, \bX) \equiv 
P(\bG_{1},\bG_{2},\bG_{3},\underline{\bY}_{1}, \underline{\bY}_{2}, \underline{\bY}_{3}, \textbf{W}, \bX)$$
We assume the joint distribution of these random variables of
all units is governed by a parameter $(\theta_X, \theta_W, \theta)$, conditional on which the random variables for each firm are i.i.d. Therefore, 
\begin{eqnarray*}
\lefteqn{P(\bG_{1},\bG_{2},\bG_{3},\underline{\bY}_{1}, \underline{\bY}_{2}, \underline{\bY}_{3}, \textbf{W}, \bX)=}\\&=&
\prod_{i=1}^{N} P(G_{i,1},G_{i,2},G_{i,3},\uY_{i,1}, \uY_{i,2}, \uY_{i,3}, W_i, \bX_i \mid  \theta) p(\theta_X, \theta_W, \theta_X, \theta_W, \theta)\\
&=&\prod_{i=1}^{N} P(\bX_i \mid  \theta_X) P(G_{i,1},G_{i,2},G_{i,3}\mid \bX_i, \theta) P(\uY_{i,1}, \uY_{i,2}, \uY_{i,3}\mid G_{i,1},G_{i,2},G_{i,3}, \bX_i, \theta) \\&&
\qquad  P(W_i\mid G_{i,1},G_{i,2},G_{i,3},\uY_{i,1}, \uY_{i,2}, \uY_{i,3}, \bX_i, \theta_W) p(\theta_X, \theta_W, \theta)
\end{eqnarray*}	
where $ p(\theta_X, \theta_W, \theta)$ is the prior distribution for $(\theta_X, \theta_W, \theta)$.
Under strong ignorability of the treatment assignment mechanism, assuming that the parameters for the models of covariates,  assignment mechanism, and   outcomes  are a priori distinct and independent, we 	have that 
\begin{eqnarray*}
\lefteqn{P(\theta \mid \bG_{1},\bG_{2},\bG_{3},\underline{\bY}_{1}, \underline{\bY}_{2}, \underline{\bY}_{3}, \textbf{W}, \bX) \propto p(\theta_X) \prod_{i=1}^{N} P(\bX_i \mid  \theta_X)\times	p(\theta_W) \prod_{i=1}^{N} P(W_i \mid \bX_i,  \theta_X) \times}\\&&
	 p(\theta) \prod_{i=1}^{N} P(G_{i,1},G_{i,2},G_{i,3}\mid \bX_i, \theta) P(\uY_{i,1}, \uY_{i,2}, \uY_{i,3}\mid G_{i,1},G_{i,2},G_{i,3}, \bX_i, \theta)
\end{eqnarray*}	
Following most of the literature, we condition on the observed values of the covariates, so we do not need to 
model the covariates, $\bX_i$. Moreover, under the above assumptions, the posterior distributions of$\theta$ and consequently of the SACEs, do not depend on the model for the assignment mechanism, which is 
is ignorable in Bayesian inference of SACEs. Therefore Bayesian inference for  SACEs requires specifying the model for the principal stratum membership,
$ P(G_{i,1},G_{i,2},G_{i,3}\mid \bX_i, \theta)$, and the outcome model, $P(\uY_{i,1}, \uY_{i,2}, \uY_{i,3}\mid G_{i,1},G_{i,2},G_{i,3}, \bX_i, \theta)$.

\subsection{Longitudinal principal stratum submodel}
We first factorize the joint distribution of the longitudinal principal stratum membership using the law of total probability as follows:
\begin{eqnarray*}
	P(G_{i1},G_{i2},G_{i3}\mid \bX_i, \theta) = 	P(G_{i1} \mid \bX_i, \theta)\times 	P(G_{i2} \mid G_{i1}, \bX_i, \theta)\times 	P(G_{i3}\mid G_{i1},G_{i2},\bX_i, \theta),
\end{eqnarray*}		
and then we model each member on the left hand using multinational logit and logit regression models (under monotonicity). Formally,
$$
P(G_{i,1}= g \mid \bX_i) = \dfrac{\exp\left\{\delta^g_{0,1} + \mathbf{\delta}^g_1\bX_i\right\} }{\sum_{ \nu \in \{AS, CS, NS\}}\exp\left\{\delta^\nu_{0,1} + \mathbf{\delta}^\nu_1\bX_i \right\}} \qquad g=AS, CS, NS.
$$
We normalize these probabilities by setting  $(\delta^{NS}_{0,1}, \mathbf{\delta}^{NS}_1)$ equal to the  vector of zeros. Then,
$$
P(G_{i,2}= g \mid G_{i,1}=AS, \bX_i) = \dfrac{\exp\left\{\delta^{g|AS}_{0,2} + \mathbf{\delta}^{g|AS}_2\bX_i \right\}}{\sum_{ \nu \in \{AS, CS, NS\}}\exp\left\{\delta^{\nu|AS}_{0,2} + \mathbf{\delta}^{\nu|AS}_2\bX_i \right\}} \qquad g=AS, CS, NS
$$
where we set $(\delta^{NS|AS}_{0,2}, \mathbf{\delta}^{NS|AS}_2)$ equal to the vector of zeros;
\begin{eqnarray*}
P(G_{i,2}= CS \mid G_{i,1}=CS, \bX_i) &=& \dfrac{\exp\left\{\delta^{CS|CS}_{0,2} + \mathbf{\delta}^{CS|CS}_2\bX_i\right\} }{1+\exp\left\{\delta^{CS|CS}_{0,2} + \mathbf{\delta}^{CS|CS}_2\bX_i\right\}}, \\ 
\\
P(G_{i,2}= NS \mid G_{i,1}=CS, \bX_i) &=& 1-\dfrac{\exp\left\{\delta^{CS|CS}_{0,2} + \mathbf{\delta}^{CS|CS}_2\bX_i\right\} }{1+\exp\left\{\delta^{CS|CS}_{0,2} + \mathbf{\delta}^{CS|CS}_2\bX_i\right\}};
\end{eqnarray*}	
and
$$
P(G_{i,2}= NA \mid G_{i,1}=NS, \bX_i) =1.
$$
Finally,
$$
P(G_{i,3}= g \mid G_{i,1}=AS, G_{i,1}=AS,\bX_i) = \dfrac{\exp\left\{\delta^{g|AS.AS}_{0,3} + \mathbf{\delta}^{g|AS.AS}_3\bX_i \right\}}{\sum_{ \nu \in \{AS, CS, NS\}}\exp\left\{\delta^{\nu|AS.AS}_{0,3} + \mathbf{\delta}^{\nu|AS.AS}_3\bX_i \right\}} \qquad g=AS, CS, NS
$$
where we set $(\delta^{NS|AS.AS}_{0,3}, \mathbf{\delta}^{NS|AS.AS}_3)$ equal to the vector of zeros;
\begin{eqnarray*}
	P(G_{i,3}= CS \mid G_{i,1}=AS,G_{i,2}= CS, \bX_i) &=& \dfrac{\exp\left\{\delta^{CS|AS.CS}_{0,3} + \mathbf{\delta}^{CS|AS.CS}_3\bX_i\right\} }{1+\exp\left\{\delta^{CS|AS.CS}_{0,3} + \mathbf{\delta}^{CS|AS.CS}_3\bX_i\right\}}, \\ 
	\\
		P(G_{i,3}= NS \mid G_{i,1}=AS,G_{i,2}= CS, \bX_i) &=&1-  \dfrac{\exp\left\{\delta^{CS|AS.CS}_{0,3} + \mathbf{\delta}^{CS|AS.CS}_3\bX_i\right\} }{1+\exp\left\{\delta^{CS|AS.CS}_{0,3} + \mathbf{\delta}^{CS|AS.CS}_3\bX_i\right\}};
\end{eqnarray*}	
\begin{eqnarray*}
	P(G_{i,3}= CS \mid G_{i,1}=CS,G_{i,2}= CS, \bX_i) &=& \dfrac{\exp\left\{\delta^{CS|CS.CS}_{0,3} + \mathbf{\delta}^{CS|CS.CS}_3\bX_i\right\} }{1+\exp\left\{\delta^{CS|AS.CS}_{0,3} + \mathbf{\delta}^{CS|CS.CS}_3\bX_i\right\}}, \\ 
	\\
	P(G_{i,3}= NS \mid G_{i,1}=CS,G_{i,2}= CS, \bX_i) &=&1- \dfrac{\exp\left\{\delta^{CS|CS.CS}_{0,3} + \mathbf{\delta}^{CS|CS.CS}_3\bX_i\right\} }{1+\exp\left\{\delta^{CS|CS.CS}_{0,3} + \mathbf{\delta}^{CS|CS.CS}_3\bX_i\right\}};
\end{eqnarray*}	
and
\begin{eqnarray*}
	P(G_{i,3}= NS \mid G_{i,1}=AS,G_{i,2}= NS, \bX_i) &=&1,\\
		P(G_{i,3}= NS \mid G_{i,1}=CS,G_{i,2}= NS, \bX_i) &=&1,\\
			P(G_{i,3}= NS \mid G_{i,1}=NS,G_{i,2}= NS, \bX_i) &=&1.
\end{eqnarray*}	

\subsection{Outcome submodels}
We assume that the unobserved potential outcome is independent of the observed potential outcome conditional on the covariates, principal stratum membership, and parameters, so that, 
$P(\uY_{i,1}, \uY_{i,2}, \uY_{i,3}\mid G_{i,1},G_{i,2},G_{i,3}, \bX_i, \theta)=
P(Y_{i,1}(0), Y_{i,2}(0), Y_{i3}(0)\mid G_{i1},G_{i2},G_{i,3}, \bX_i, \theta)\times P(Y_{i,1}(1), Y_{i,2}(1), Y_{i,3}(1)\mid G_{i,1},G_{i,2},G_{i,3}, \bX_i, \theta)
$
and we only need to specify the marginal distribution for the control and treatment potential outcomes.
 For $w=0,1$, 
\begin{eqnarray*}
\lefteqn{P(Y_{i,1}(w), Y_{i,2}(w), Y_{i,3}(w)\mid G_{i,1},G_{i,2},G_{i3}, \bX_i, \theta)=}\\&&
P(Y_{i,1}(w) \mid G_{i,1},G_{i,2},G_{i3}, \bX_i, \theta)\times
P(Y_{i,2}(w) \mid Y_{i,1}(w), G_{i,1},G_{i,2},G_{i3}, \bX_i, \theta)\times\\&&
P(Y_{i,3}(w) \mid Y_{i,1}(w), Y_{i,2}(w), G_{i,1},G_{i,2},G_{i3}, \bX_i, \theta)
\end{eqnarray*}	
We assume that the hiring decision at time $t$ is not affected by future principal stratum membership conditionally on the current and past principal stratum membership, past hiring decision, covariates, and parameters. 
\begin{assumption}\label{ass:expectations} For all $i$
		$$P(Y_{i,1}(w)\mid G_{i,1}, G_{i,2}, G_{i,3},\bX_i, \theta) = P(Y_{i,1}(w)\mid G_{i,1}, \bX_i, \theta)  
		$$
		and 
		$$P(Y_{i,2}(w)\mid Y_{i,1}(w), G_{i,1}, G_{i,2}, G_{i,3},\bX_i,  \theta) = P(Y_{i,2}(w)\mid Y_{i,1}(w), G_{i,1},  G_{i,2}, \bX_i,  \theta) 
		$$
		for $w=0,1$.
	\end{assumption}
In our study, Assumption~\ref{ass:expectations} is reasonable. Underlying this assumption is the notion that present hiring is ruled by the present ``health status'' and needs of the firm. 
We also make a Markov assumption, implying that the conditional probability distribution of the hiring decision at time $t=3$ is independent of  
the hiring decision at time $t=1$, conditionally on the  principal stratum membership,   hiring decision at time $t=3$, covariates and parameters:
\begin{assumption}\label{ass:markoc} For all $i$
	$$P(Y_{i,3}(w)\mid Y_{i,1}(w),Y_{i,2}(w), G_{i,1}, G_{i,2}, G_{i,3},\bX_i,  \theta) = P(Y_{i,3}(w)\mid  Y_{i,2}(w), G_{i,1}, G_{i,2}, G_{i,3},\bX_i,  \theta)
	$$
	for $w=0,1$.
\end{assumption}
Because our outcome, hiring decision,  is dichotomous for the subpopulations where it is defined, we assume that outcome distributions take the form of logistic regressions. Specifically, we need to specify twelve logit regression models. 
\begin{eqnarray*}
P(Y_{i,1}(w)=1\mid G_{i,1}=AS, \bX_i, \theta)  &=&
\dfrac{\exp\left\{\beta^{w,AS}_{0,1} + \mathbf{\beta}^{w,AS}_1\bX_i\right\} }{1+\exp\left\{\beta^{w,AS}_{0,1} + \mathbf{\beta}^{w,AS}_1\bX_i\right\} }, 
\qquad w=0,1;\\
\\
	P(Y_{i,1}(1)=1\mid G_{i,1}=CS, \bX_i, \theta) &=&
	\dfrac{\exp\left\{\beta^{1,CS}_{0,1} + \mathbf{\beta}^{1,CS}_1\bX_i\right\} }{1+\exp\left\{\beta^{1,CS}_{0,1} + \mathbf{\beta}^{1,CS}_1\bX_i\right\} }  
\end{eqnarray*}

\begin{eqnarray*}
\lefteqn{ P(Y_{i,2}(w)=1\mid Y_{i,1}(w), G_{i,1}=AS,  G_{i,2}=AS, \bX_i,  \theta)=}\\&& \qquad \qquad
\dfrac{\exp\left\{\beta^{w,AS.AS}_{0,2} +\lambda^{w,AS.AS}_{2}Y_{i,1}(w)+ \mathbf{\beta}^{w,AS.AS}_2\bX_i\right\} }{1+\exp\left\{\beta^{wAS.AS}_{0,2} +
	\lambda^{w,AS.AS}_{2}Y_{i,1}(w)+
	\mathbf{\beta}^{w,AS.AS}_2\bX_i\right\} }, \qquad w=0,1;\\
\\
\lefteqn{ P(Y_{i,2}(1)=1\mid Y_{i,1}(1), G_{i,1}=g_1,  G_{i,2}=g_2, \bX_i,  \theta) =}\\&& 
\dfrac{\exp\left\{\beta^{1,g_1.g_2}_{0,2} +\lambda^{1,g_1.g_2}_{2}Y_{i,1}(1)+ \mathbf{\beta}^{1,g_1.g_2}_2\bX_i\right\} }{1+\exp\left\{\beta^{1,g_1.g_2}_{0,2} +
	\lambda^{1,g_1.g_2}_{2}Y_{i,1}(1)+
	\mathbf{\beta}^{1,g_1.g_2}_2\bX_i\right\} }, \qquad g_1.g_2 \in \{AS.CS,   CS.CS\};
\end{eqnarray*}

\begin{eqnarray*}
	\lefteqn{P(Y_{i,3}(w)=1\mid Y_{i,1}(w), G_{i,1}=AS,  G_{i,2}=AS, G_{i,3}=AS,\bX_i,  \theta)=
 }\\&& \qquad \qquad
\dfrac{\exp\left\{\beta^{w,AS.AS.AS}_{0,3} +\lambda^{w,AS.AS.AS}_{3}Y_{i,2}(w)+ \mathbf{\beta}^{w,AS.AS.AS}_3\bX_i\right\} }{1+\exp\left\{\beta^{w,AS.AS.AS}_{0,3} +\lambda^{w,AS.AS.AS}_{3}Y_{i,2}(w)+ \mathbf{\beta}^{w,AS.AS.AS}_3\bX_i\right\} }, \qquad w=0,1;\\
\\
\lefteqn{ P(Y_{i,3}(1)=1\mid Y_{i,2}(1), G_{i,1}=g_1,  G_{i,2}=g_2, G_{i,3}=g_3,\bX_i,  \theta) =}\\&& 
\dfrac{\exp\left\{\beta^{1,g_1.g_2.g_3}_{0,3} +\lambda^{1,g_1.g_2.g_3}_{3}Y_{i,2}(1)+ \mathbf{\beta}^{1,g_1.g_2.g_3}_3\bX_i\right\} }{1+\exp\left\{\beta^{1,g_1.g_2.g_3}_{0,3} +
	\lambda^{1,g_1.g_2.g_3}_{3}Y_{i,2}(1)+
	\mathbf{\beta}^{1,g_1.g_2.g_3}_3\bX_i\right\} }, \quad g_1.g_2.g_3 \in \{AS.AS.CS,   AS.CS.CS, CS.CS.CS\}.
\end{eqnarray*}
In the application in this paper, we impose prior equality of the  coefficients of the covariates in the outcome regressions: 
$\beta_1^{0,AS}=\beta_1^{1,AS}=\beta_1^{1,CS}$ $=$ 
$\beta_2^{0,AS.AS}=\beta_2^{1,AS.AS}=\beta_2^{1,AS.CS}=\beta_2^{1,CS.CS}$ 
$=$ 
$\beta_3^{0,AS.AS.AS}=\beta_3^{1,AS.AS.AS}=\beta_3^{1,AS.AS.CS}=
\beta_3^{1,AS.CS.CS}=\beta_3^{1,CS.CS.CS} \equiv \beta_X$.

We specify proper, yet weakly informative, Normal prior distributions (see web appendix for details). 
We derive the posterior distribution of the model parameters and the causal estimands of interest by implementing a Hamiltonian Monte Carlo (HMC) algorithm in RStan. We use 2000 iterations with 1000 warm-up iterations. No pathological behavior is found in the diagnostic results. We also conduct posterior predictive checks, finding no evidence against our modeling assumptions. Details are shown in the web appendix.

\FloatBarrier

\section{Results}\label{sec:results}
\subsection{Principal strata membership}

 Table \ref{tab:long_strata} shows summary statistics for the probabilities of longitudinal principal stratum membership. We note that the majority of firms are compliant survivors in each time point (58.7\%) that are carrying on their activity because of the bank loan. A sizeable part of firms are classified as always survivors in each time point (34.5\%) suggesting that around one-third of the units exhibit some level of entrepreneurial ability and their businesses would have survived irrespective of treatment assignment. Few units, 2.4\%, would have failed anyway. 
The strata, $AS.CS.NS$, $AS.AS.CS$, and $CS.NS.NS$ are very uncommon, as the posterior mean of belonging to these strata is lower than 1\%. 
 It is worth noting that transitions between different categories, such as from always survivors to compliant survivors are also relatively uncommon.

 These results confirm the rationale of public support for start-ups: easing credit access is a key factor for the survival of these firms. Even if this is a positive outcome,
 it is worth noting that most of these firms would not have survived in the market without the loan, which in turn suggests that their intrinsic entrepreneurial potential is not very high.
This raises the question of whether this type of support is the most effective use of public funds, and whether it exposes - especially when involving never survivors -  public actors to the financial risk of unreturned loans.
Figure \ref{fig:cov_strata} in the Appendix reports the boxplots for each covariate across different longitudinal principal strata.
Compliant survivors are more likely to have a young owner and to be new establishments in the retail and service sectors. On the other hand, always survivors are often more established businesses, owned by women, and operating in the manufacturing sector, which is relatively rare among compliant survivors. Compliant survivors also tend to rely more on local bank loans, while always survivors may have access to national banks due to their stronger financial standing. In general, there seems to be a distinction between established firms with higher potential and lower-potential start-up projects that are mainly led by young entrepreneurs and are heavily reliant on bank loans in the first period.

\begin{table}[t]
\caption{Summary statistics of the posterior probability of longitudinal principal strata memberships}
\centering
\label{tab:long_strata}
\begin{tabular}[t]{l|c|c|c|c}
\hline 
 $\textbf{G}_i$  & \textbf{Mean}& \textbf{st.dev}  &\textbf{ 0.05} &\textbf{ 0.95} \\
\hline
$NS.NS.NS$ & 0.024 & 0.002 &  0.021 & 0.027 \\
$CS.NS.NS$ & 0.007 & 0.002 &  0.003 & 0.011 \\
$AS.NS.NS$ & 0.046 & 0.003  & 0.041 & 0.051 \\
$AS.AS.NS$ & 0.016 & 0.007  & 0.007 & 0.031 \\
$AS.CS.NS$ & 0.000 & 0.000  & 0.000 & 0.001 \\
$CS.CS.NS$ & 0.037 & 0.007  & 0.022 & 0.047 \\
$CS.CS.CS$ & 0.587 & 0.138  & 0.322 & 0.778 \\
$AS.CS.CS$ & 0.013 & 0.036  & 0.000 & 0.070 \\
$AS.AS.CS$ & 0.002 & 0.003  & 0.000 & 0.009 \\
$AS.AS.AS$ & 0.345 & 0.130  & 0.165 & 0.587 \\
\hline
\end{tabular}
\end{table}

\subsection{Principal strata effects}

\begin{table}[ht]
    \centering
       \caption{Summary statistics of the posterior distribution of SACEs}
    \begin{tabular}{cccccc}
    \hline 
       & \textbf{Mean} & \textbf{st.dev} & \textbf{0.05} & \textbf{0.95} \\
        \hline
        $SACE_1(1)$ & 0.102 & 0.057 & 0.008 & 0.197 \\
        $SACE_{1:2}(2)$  & 0.054 & 0.068 & -0.051 & 0.175\\
        $SACE_{1:3}(3)$ & -0.028 & 0.055 & -0.109 & 0.076 \\
        \hline
    \end{tabular}
 
    \label{tab:SACE}
\end{table}

We now focus on the subpopulations of always survivors for which we can evaluate the causal effects of the bank loan on hiring decisions.
The posterior means (standard deviations) of the probabilities that a firm is an always survivor at time $t=1$, $t=1,2$, and $t=1,2,3$ are, respectively, 0.384(0.129), 0.345(0.128) and 0.345(0.130). 

\begin{figure}[ht]
    \centering
    \scalebox{0.95}{
    \includegraphics[width=\textwidth]{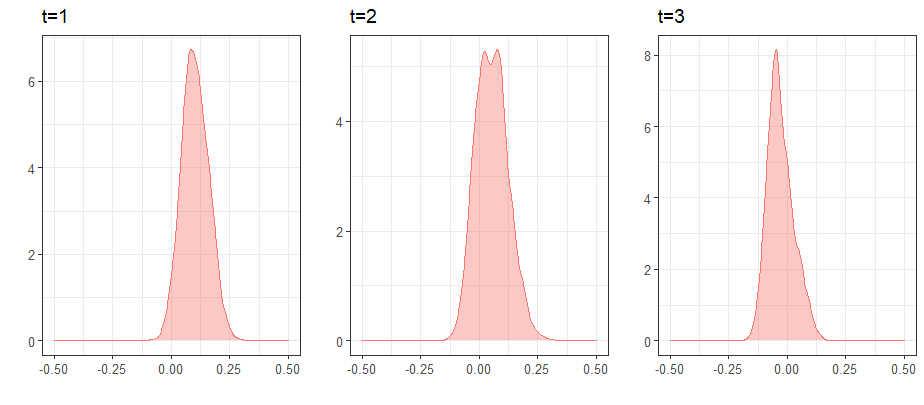}}
    \caption{Posterior distribution of $SACE_1(1)$, $SACE_{1:2}(2)$ and $SACE_{1:3}(3)$}
    \label{fig:SACE}
\end{figure}

Table \ref{tab:SACE}  shows summary statistics (mean, standard deviation, and fifth and ninety-fifth quantiles) of the posterior distributions of the SACEs, and Figure \ref{fig:SACE} shows their posterior distributions.
We can get precious insights from the analysis of such effects. 
We find a positive effect of the treatment on firms' hiring decisions in the first period after the bank loan, $SACE_1(1)$. We can hypothesize that the additional liquidity provided by the bank loan helps firms not only to start the activity but also to hire the human resources needed for the firm's operations. Instead, for firms established recently, it is possible that this loan supports an enlargement opportunity with an increase in the number of employees. The causal effect $SACE_{1:2}(2)$  is also positive, but small and not statistically significant. $SACE_{1:3}(3)$ is very small and   statistically negligible.

\begin{figure}[H]
    \centering
    \scalebox{0.95}{
    \includegraphics[width=\textwidth]{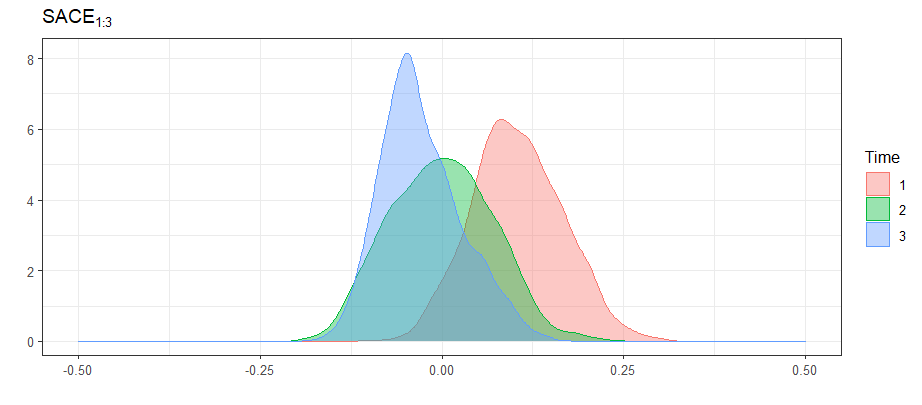}}
    \caption{Posterior distributions of $SACE_{1:3}$ for $t \in \{1,2,3\}$ with $G_i = AS.AS.AS$}
    \label{fig:long_sace}
\end{figure}

Table \ref{tab:long_sace} and Figure \ref{fig:long_sace} show the posterior distributions of $SACE_{1:3}(t)$, for $t=1,2,3$, which are causal effects for those units that are classified as always survivor in all three times. 
This is the only principal strata for which the hiring decision is defined for all post-treatment periods and levels of treatment, and thus we can study the evolution of the effects over time. 

The estimated SACE effects decrease with time. $SACE_{1:3}(t)$ is positive and statistically significant in the first period. During the second period, the effect is still positive, but smaller and not statistically significant. In the last period, the effect is very small and not statistically significant. 
These results can be reasonable: firms hire new workers during the kick-off periods of the financed project, and these projects seem to have not generated enough growth to allow for additional hiring in the subsequent years. This is a bit of a disappointing result because even the more promising projects such as those led by always survivors have not generated enough sales to justify further investment in human resources.
The longitudinal $\text{SACE}_{1:2}$ is reported in the appendix and confirms the results obtained for $\text{SACE}_{1:3}$.

\begin{table}
    \centering
    \begin{tabular}{ccccc}
        \hline
        & \textbf{Mean} & \textbf{st.dev} & \textbf{0.05} & \textbf{0.95} \\
        \hline
      $SACE_{1:3}(1)$ & 0.104 & 0.062 & 0.002 & 0.210 \\
       $SACE_{1:3}(2)$ & 0.008 & 0.072 & -0.108 & 0.129 \\
      $SACE_{1:3}(3)$  & -0.028 & 0.055 & -0.109 & 0.076 \\
                 \hline
    \end{tabular}
    \caption{Longitudinal SACE effects for AS.AS.AS units}
    \label{tab:long_sace}
\end{table}

\FloatBarrier

\section{Conclusions} \label{sec:conclusion}

In this study, we propose a framework for studying the causal effects of a policy in the presence of non-ignorable censoring and truncation by death. 
In our motivating application, it is impossible to evaluate directly the policy effects on the main outcome (hiring decisions made by the funded start-ups) after the outcome was censored by the closure of the business.

To address this problem, we follow a principal stratification approach, under the potential outcomes framework. Expanding on \cite{bia2020}, we modify principal stratification to cope with the longitudinal structure of our data, with multiple post-treatment periods. We identify principal strata based on treatment status and the survival of the firms in each post-treatment period.
This allows us to identify the principal strata that would have survived regardless of their treatment assignment and to estimate the causal effects within these strata.

We exploit a Bayesian approach for inference and impute the missing potential outcomes using a data augmentation algorithm that we use to derive the posterior distribution for longitudinal stratum membership and causal effects on hiring decisions.

Results from policy evaluations of start-up programs have raised debate in the economic policy literature. Few studies focus on the second-round effect of such policies, the so-called double dividend, and even fewer of these exploit a proper causal structure. In this work, we explicitly address both of these gaps. We focus on the secondary outcome of this policy, which is the hiring decisions made by funded start-ups, and aim to determine if this policy has the potential to provide a "double dividend" for the community, beyond the self-employment prospects of the entrepreneur.

Our results highlight that there may be relevant differences among entrepreneurial projects: the majority of new firms are heavily dependent on public support for their survival and they would have stopped their activity under the control assignment. On the other hand, about 35 \% of the firms would have remained active in the market regardless of their treatment assignment, suggesting that they have higher potential.

Causal effects on hiring decisions are accessible only for firms who would have survived whatever their treatment assignment.  
We find positive effects from the policy on the stratum of always survivors, with decreasing amounts through consecutive times. This suggests that the policy is successful in encouraging hiring decisions by self-employed entrepreneurs, but once these new positions are filled, the start-ups struggle to expand further.

Our research concurs with the conclusions of \cite{mariani2019} in suggesting that the provision of public support for facilitating credit market access constitutes a viable active labor market strategy that combats unemployment through self-employment. However, it must be acknowledged that such a strategy entails the use of public resources in favor of projects whose low potential may be open to debate.

 \bibliographystyle{apalike}
\bibliography{bibtex}

\include{app4.tex}

\end{document}

%% file: app4.tex
\section{Appendix}

\FloatBarrier
\subsection{Additional Results}

\begin{table}
    \centering
    \begin{tabular}{ccccc}
        \hline
        & \textbf{Mean} & \textbf{st.dev} & \textbf{0.05} & \textbf{0.95} \\
        \hline
      $SACE_{1:3}(1)$ & 0.102 & 0.062 & 0.003 & 0.205 \\
       $SACE_{1:3}(2)$ & 0.054 & 0.068 & -0.108 & 0.129 \\
     \hline
    \end{tabular}
    \caption{Longitudinal SACE effects for AS.AS.AS units}
    \label{tab:long_sace2}
\end{table}

\begin{figure}[H]
    \centering
    \scalebox{0.95}{
    \includegraphics[width=\textwidth]{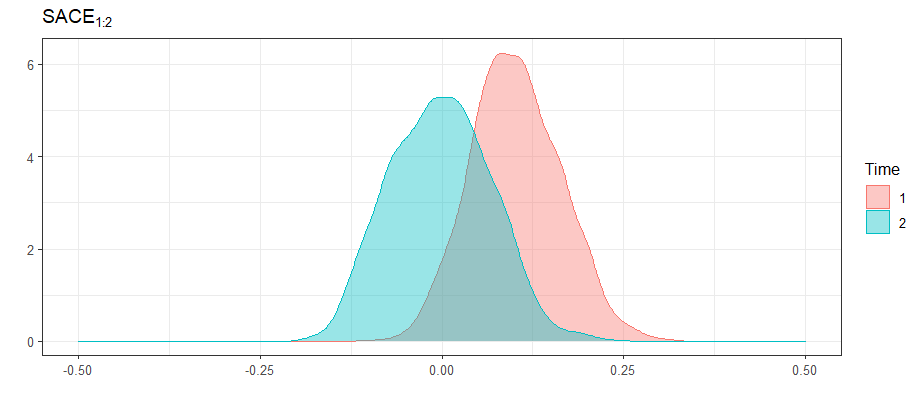}}
    \caption{Posterior distributions of $SACE_{1:2}$ for $t \in \{1,2\}$ with $G_i = AS.AS$}
    \label{fig:long_sace2}
\end{figure}

\FloatBarrier
\subsection{Priors specification}

In this subsection of the appendix we report the prior specifications we used in our analysis. We assume prior distributions are a-prior independent. 


\begin{itemize}
    \item Priors for the strata membership model
\end{itemize}

\begin{equation*}\label{eq:priors_strata}
\begin{aligned}
\delta^g_{0,1}= \delta^{g|AS}_{0,2}=\delta^{g|AS.AS}_{0,3} \sim \mathcal{N}(0,2.5)\\
\delta^g_{1}= \delta^{g|AS}_{2}=\delta^{g|AS.AS}_{3} \sim \mathcal{N}(0,2.5)\\
\delta^{\nu}_{0,1}= \delta^{\nu|AS}_{0,2}=\delta^{\nu|AS.AS}_{0,3} \sim \mathcal{N}(0,2.5)\\
\delta^{\nu}_{1}= \delta^{\nu|AS}_{2}=\delta^{\nu|AS.AS}_{3} \sim \mathcal{N}(0,2.5)\\
\delta^{CS|CS}_{0,2}= \delta^{CS|CS}_{2} \sim \mathcal{N}(0,2.5)\\
\delta^{CS|AS.CS}_{0,3}= \delta^{CS|AS.CS}_{3} \sim \mathcal{N}(0,2.5)\\
\delta^{CS|CS.CS}_{0,3}= \delta^{CS|CS.CS}_{3} \sim \mathcal{N}(0,2.5)\\
\end{aligned}
\end{equation*}

\begin{itemize}
    \item Priors for the strata membership model
\end{itemize}

\begin{equation*}
\label{eq:priors_outcome}
\begin{aligned}
\beta^{w, AS}_{0,1}= \beta^{w,AS.AS}_{0,2}=\beta^{w,AS.AS.AS}_{0,3} \sim \mathcal{N}(0,2)\\
\beta^{w,AS}_{1}= \beta^{w,AS.AS}_{2}=\beta^{w,AS.AS.AS}_{3} \sim \mathcal{N}(0,2)\\
\beta^{1, CS}_{0,1}= \beta^{1,CS}_{1} \sim \mathcal{N}(0,2)\\
\beta^{1,g1.g2}_{0,2}=\beta^{1,g1.g2.g3}_{0,3} \sim \mathcal{N}(0,2)\\
\beta^{1,g1.g2}_{2}=\beta^{1,g1.g2.g3}_{3} \sim \mathcal{N}(0,2)\\
\lambda^{1,g1.g2}_{0,2}=\lambda^{1,g1.g2.g3}_{0,3} \sim \mathcal{N}(0,2)\\
\lambda^{1,g1.g2}_{2}=\lambda^{1,g1.g2.g3}_{3} \sim \mathcal{N}(0,2)\\
\lambda^{w,AS.AS}_{0,2}=\lambda^{w,AS.AS.AS}_{0,3} \sim \mathcal{N}(0,2)\\
\lambda^{w,AS.AS}_{2}=\lambda^{w,AS.AS.AS}_{3} \sim \mathcal{N}(0,2)\\
\end{aligned}
\end{equation*}

\FloatBarrier
\subsection{Figures}


\begin{figure}
    \centering
    \includegraphics[width=0.5\textwidth]{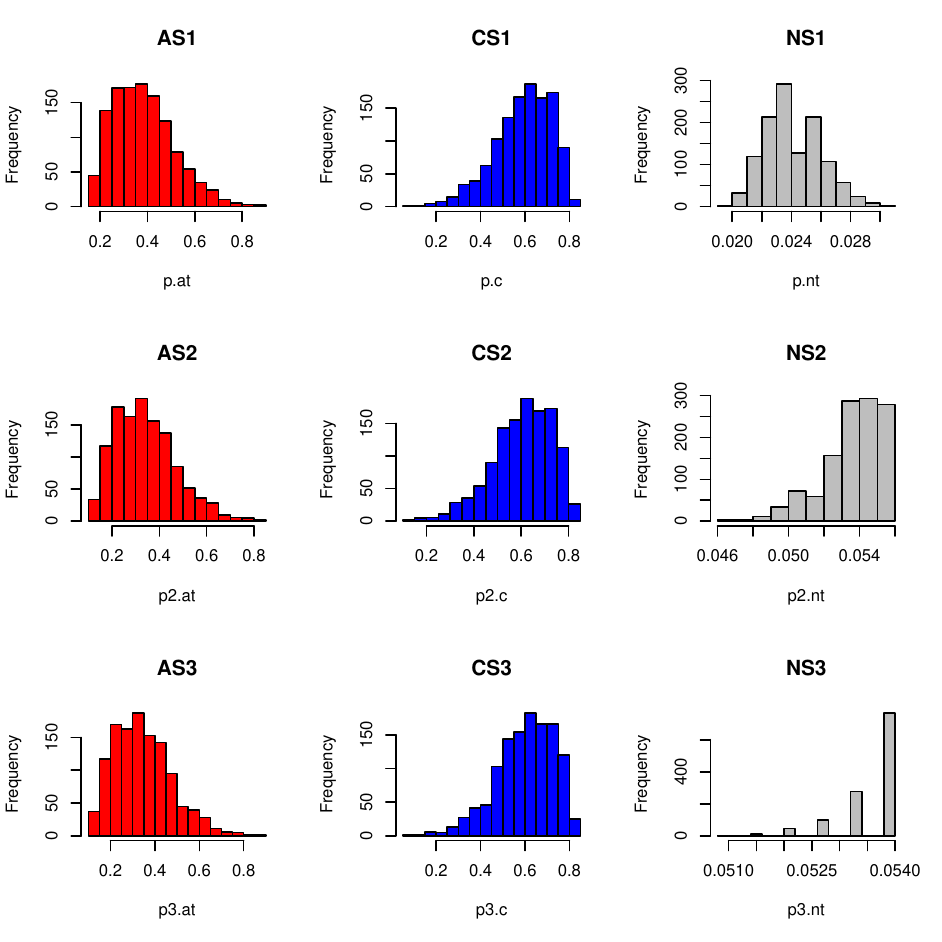}
    \caption{Posterior probability for principal strata membership in each post-treatment period}
    \label{fig:strata_hist}
\end{figure}

\begin{figure}
    \centering
    \includegraphics[width=.8\textwidth]{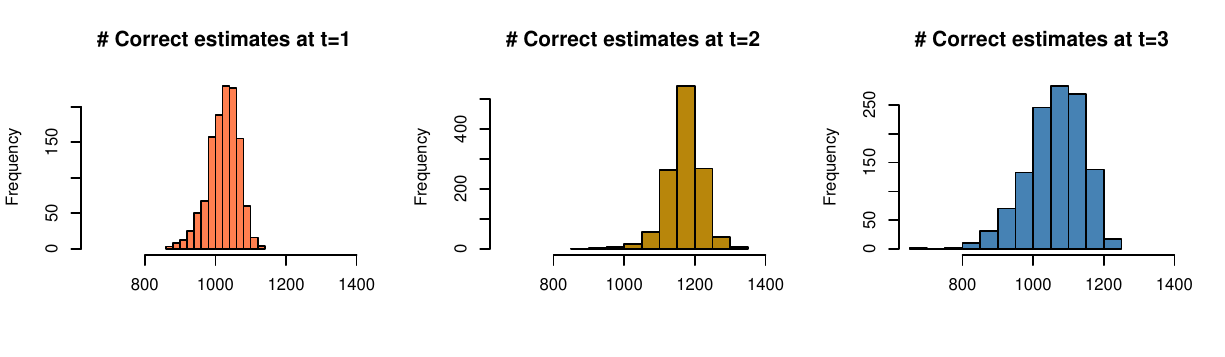}
    \caption{Number of corrected estimated outcomes for hiring decision in each period $t \in {1,2,3}$ over the HMC iterations}
    \label{fig:my_label}
\end{figure}

\begin{figure}
    \centering
    \includegraphics[width=.8\textwidth]{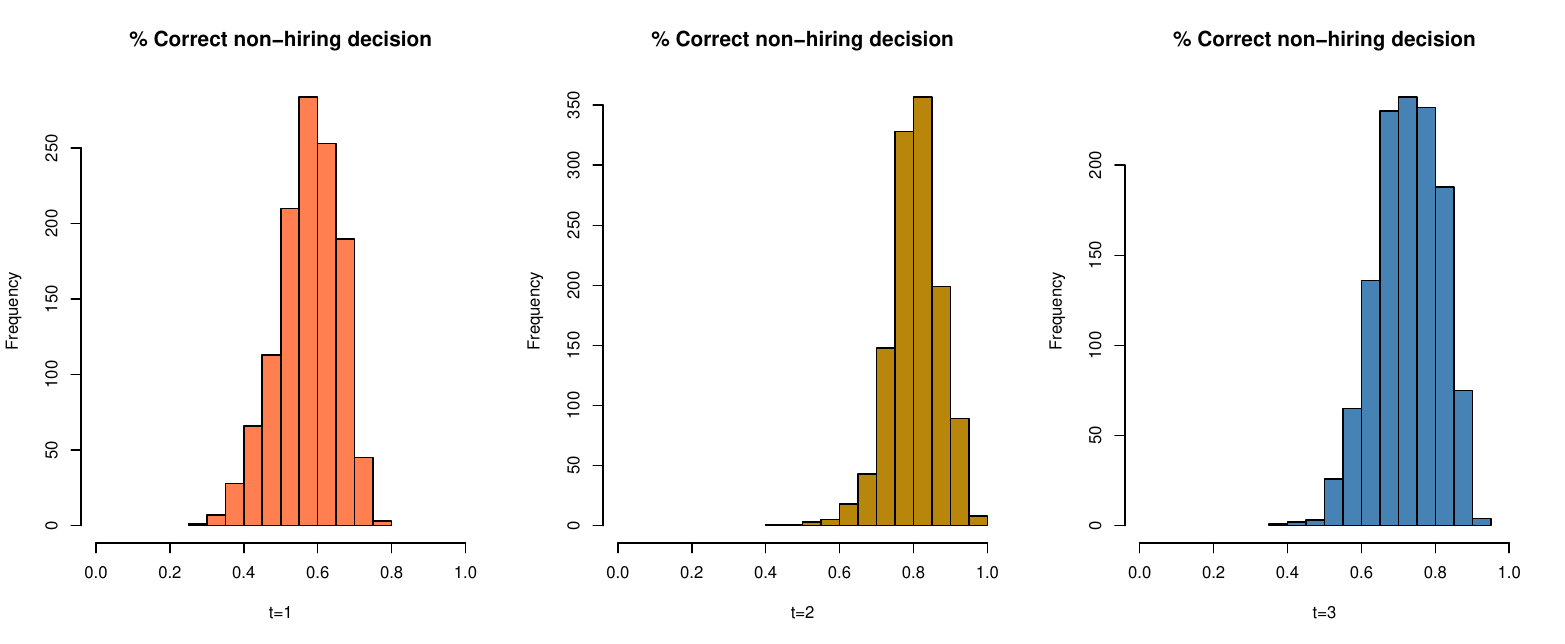}
    \caption{\% of correct predictions for non-hiring decisions}
    \label{fig:sens}
\end{figure}

\begin{figure}
    \centering
    \includegraphics[width=.8\textwidth]{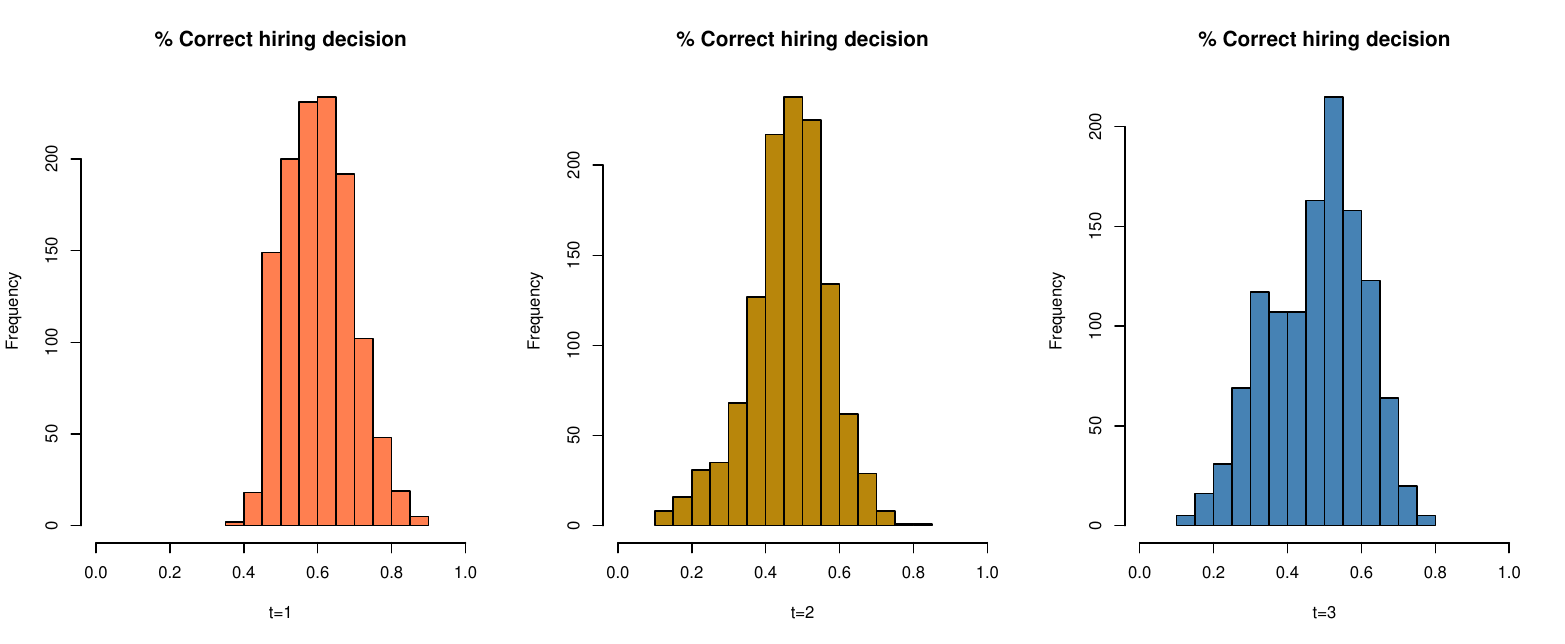}
    \caption{\% of correct predictions for hiring decisions}
    \label{fig:spec}
\end{figure}

\begin{landscape}
\begin{figure}
    \centering
    \includegraphics[width=1.5\textwidth]{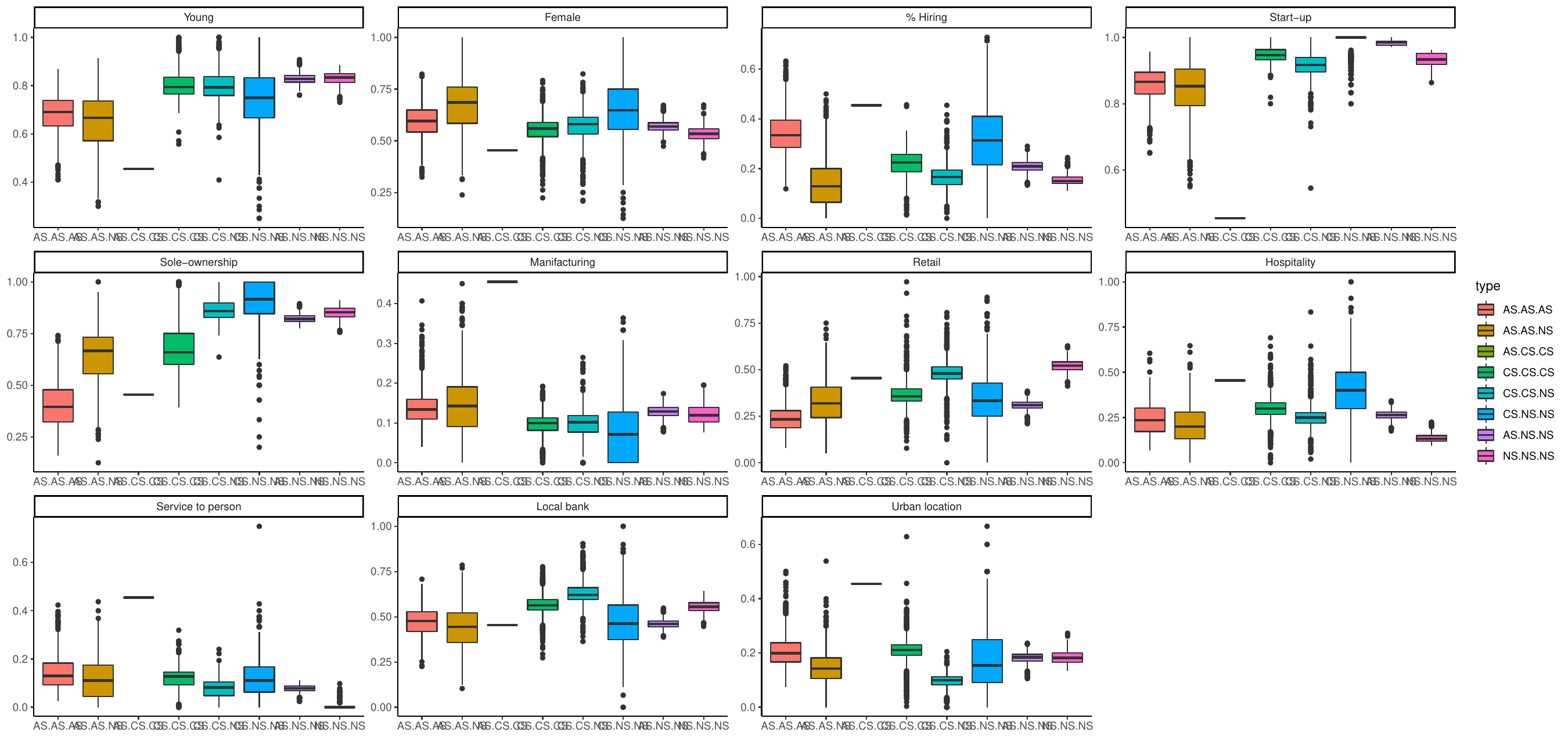}
    \caption{Boxplots of covariates across the longitudinal strata}
    \label{fig:cov_strata}
\end{figure}
\end{landscape}

%% file: ECOSTA.bbl
\begin{thebibliography}{}

\bibitem[Antonelli et~al., 2023]{antonelli2023principal}
Antonelli, J., Mealli, F., Beck, B., and Mattei, A. (2023).
\newblock Principal stratification with continuous treatments and continuous
  post-treatment variables.
\newblock {\em arXiv preprint arXiv:2309.14486}.

\bibitem[Battistin et~al., 2001]{battistin2001subsidised}
Battistin, E., Gavosto, A., and Rettore, E. (2001).
\newblock Why do subsidised firms survive longer? an evaluation of a program
  promoting youth entrepreneurship in italy.
\newblock In {\em Econometric evaluation of labour market policies}, pages
  153--181. Springer.

\bibitem[Bia et~al., 2020]{bia2020}
Bia, M., Mattei, A., and Mercatanti, A. (2020).
\newblock Assessing causal effects in a longitudinal observational study with
  “truncated” outcomes due to unemployment and nonignorable missing data.
\newblock {\em Journal of Business \& Economic Statistics}, pages 1--34.

\bibitem[Caliendo, 2016]{caliendo2016}
Caliendo, M. (2016).
\newblock Start-up subsidies for the unemployed: Opportunities and limitations.
\newblock {\em IZA World of Labor}, 200:10.15185/izawol.200.

\bibitem[Caliendo et~al., 2015]{caliendo2015subsidized}
Caliendo, M., Hogenacker, J., K{\"u}nn, S., and Wie{\ss}ner, F. (2015).
\newblock Subsidized start-ups out of unemployment: a comparison to regular
  business start-ups.
\newblock {\em Small Business Economics}, 45:165--190.

\bibitem[Caliendo and K{\"u}nn, 2014]{caliendo2014regional}
Caliendo, M. and K{\"u}nn, S. (2014).
\newblock Regional effect heterogeneity of start-up subsidies for the
  unemployed.
\newblock {\em Regional Studies}, 48(6):1108--1134.

\bibitem[Chiba and VanderWeele, 2011]{chiba2011simple}
Chiba, Y. and VanderWeele, T.~J. (2011).
\newblock A simple method for principal strata effects when the outcome has
  been truncated due to death.
\newblock {\em American journal of epidemiology}, 173(7):745--751.

\bibitem[Colombo and Grilli, 2007]{colombo2007funding}
Colombo, M.~G. and Grilli, L. (2007).
\newblock Funding gaps? access to bank loans by high-tech start-ups.
\newblock {\em Small Business Economics}, 29(1):25--46.

\bibitem[Decker and Corson, 1995]{decker1995international}
Decker, P.~T. and Corson, W. (1995).
\newblock International trade and worker displacement: Evaluation of the trade
  adjustment assistance program.
\newblock {\em ILR Review}, 48(4):758--774.

\bibitem[Decker et~al., 2014]{decker2014role}
Decker, R., Haltiwanger, J., Jarmin, R., and Miranda, J. (2014).
\newblock The role of entrepreneurship in us job creation and economic
  dynamism.
\newblock {\em Journal of Economic Perspectives}, 28(3):3--24.

\bibitem[Dumont et~al., 2016]{dumont2016contribution}
Dumont, M., Rayp, G., Verschelde, M., and Merlevede, B. (2016).
\newblock The contribution of start-ups and young firms to industry-level
  efficiency growth.
\newblock {\em Applied Economics}, 48(59):5786--5801.

\bibitem[Dvouletý and Lukeš, 2016]{dvoulety2016}
Dvouletý, O. and Lukeš, M. (2016).
\newblock Review of empirical studies on self-employment out of unemployment:
  Do self-employment policies make a positive impact?
\newblock {\em International Review of Entrepreneurship}, 14(3):361--376.

\bibitem[Fairclough et~al., 1998]{fairclough1998missing}
Fairclough, D.~L., Peterson, H.~F., and Chang, V. (1998).
\newblock Why are missing quality of life data a problem in clinical trials of
  cancer therapy?
\newblock {\em Statistics in medicine}, 17(5-7):667--677.

\bibitem[Forastiere et~al., 2016]{forastiere2016identification}
Forastiere, L., Mealli, F., and VanderWeele, T.~J. (2016).
\newblock Identification and estimation of causal mechanisms in clustered
  encouragement designs: Disentangling bed nets using bayesian principal
  stratification.
\newblock {\em Journal of the American Statistical Association},
  111(514):510--525.

\bibitem[Frangakis and Rubin, 2002]{frangakis2002}
Frangakis, C.~E. and Rubin, D.~B. (2002).
\newblock Principal stratification in causal inference.
\newblock {\em Biometrics}, 58(1):21--29.

\bibitem[Frumento et~al., 2012a]{frumento2012evaluating}
Frumento, P., Mealli, F., Pacini, B., and Rubin, D.~B. (2012a).
\newblock Evaluating the effect of training on wages in the presence of
  noncompliance, nonemployment, and missing outcome data.
\newblock {\em Journal of the American Statistical Association},
  107(498):450--466.

\bibitem[Frumento et~al., 2012b]{frumento2012}
Frumento, P., Mealli, F., Pacini, B., and Rubin, D.~B. (2012b).
\newblock Evaluating the effect of training on wages in the presence of
  noncompliance, nonemployment, and missing outcome data.
\newblock {\em Journal of the American Statistical Association},
  107(498):450--466.

\bibitem[Gustafson, 2010]{gustafson2010bayesian}
Gustafson, P. (2010).
\newblock Bayesian inference for partially identified models.
\newblock {\em The international journal of biostatistics}, 6(2).

\bibitem[Haltiwanger et~al., 2013]{haltiwanger2013creates}
Haltiwanger, J., Jarmin, R.~S., and Miranda, J. (2013).
\newblock Who creates jobs? small versus large versus young.
\newblock {\em Review of Economics and Statistics}, 95(2):347--361.

\bibitem[Heckman, 1976]{heckman1976common}
Heckman, J.~J. (1976).
\newblock The common structure of statistical models of truncation, sample
  selection and limited dependent variables and a simple estimator for such
  models.
\newblock In {\em Annals of economic and social measurement, volume 5, number
  4}, pages 475--492. NBER.

\bibitem[Heckman, 1979]{heckman1979sample}
Heckman, J.~J. (1979).
\newblock Sample selection bias as a specification error.
\newblock {\em Econometrica: Journal of the econometric society}, pages
  153--161.

\bibitem[Imbens and Rubin, 1997]{imbens1997}
Imbens, G.~W. and Rubin, D.~B. (1997).
\newblock Bayesian inference for causal effects in randomized experiments with
  noncompliance.
\newblock {\em The annals of statistics}, pages 305--327.

\bibitem[Kane, 2010]{kane2010importance}
Kane, T.~J. (2010).
\newblock The importance of startups in job creation and job destruction.
\newblock {\em Available at SSRN 1646934}.

\bibitem[Li et~al., 2023]{li2023bayesian}
Li, F., Ding, P., and Mealli, F. (2023).
\newblock Bayesian causal inference: a critical review.
\newblock {\em Philosophical Transactions of the Royal Society A},
  381(2247):20220153.

\bibitem[Lin et~al., 1996]{lin1996comparing}
Lin, D., Robins, J., and Wei, L. (1996).
\newblock Comparing two failure time distributions in the presence of dependent
  censoring.
\newblock {\em Biometrika}, 83(2):381--393.

\bibitem[Luke{\v{s}} et~al., 2019]{lukevs2019business}
Luke{\v{s}}, M., Longo, M.~C., and Zouhar, J. (2019).
\newblock Do business incubators really enhance entrepreneurial growth?
  evidence from a large sample of innovative italian start-ups.
\newblock {\em Technovation}, 82:25--34.

\bibitem[Manaresi et~al., 2021]{manaresi2021supporting}
Manaresi, F., Menon, C., and Santoleri, P. (2021).
\newblock Supporting innovative entrepreneurship: an evaluation of the italian
  “start-up act”.
\newblock {\em Industrial and Corporate Change}, 30(6):1591--1614.

\bibitem[Mariani et~al., 2019]{mariani2019}
Mariani, M., Mattei, A., Storchi, L., and Vignoli, D. (2019).
\newblock The ambiguous effects of public assistance to youth and female
  start-ups between job creation and entrepreneurship enhancement.
\newblock {\em Scienze Regionali}, 18(2):237--260.

\bibitem[Mattei and Mealli, 2007]{mattei2007}
Mattei, A. and Mealli, F. (2007).
\newblock Application of the principal stratification approach to the faenza
  randomized experiment on breast self-examination.
\newblock {\em Biometrics}, 63(2):437--446.

\bibitem[Mattei and Mealli, 2011]{mattei2011}
Mattei, A. and Mealli, F. (2011).
\newblock Augmented designs to assess principal strata direct effects.
\newblock {\em Journal of the Royal Statistical Society: Series B (Statistical
  Methodology)}, 73(5):729--752.

\bibitem[McMahon and Harrell~Jr, 2001]{mcmahon2001joint}
McMahon, R.~P. and Harrell~Jr, F.~E. (2001).
\newblock Joint testing of mortality and a non-fatal outcome in clinical
  trials.
\newblock {\em Statistics in medicine}, 20(8):1165--1172.

\bibitem[Mealli and Mattei, 2012]{mealli2012}
Mealli, F. and Mattei, A. (2012).
\newblock A refreshing account of principal stratification.
\newblock {\em The international journal of biostatistics}, 8(1).

\bibitem[Mealli and Pagni, 2001]{mealli2001analisi}
Mealli, F. and Pagni, R. (2001).
\newblock Analisi e valutazione delle politiche per le nuove imprese.
\newblock {\em Il caso della LR Toscana}, 1(27/93).

\bibitem[Nigam et~al., 2020]{nigam2020impact}
Nigam, N., Mbarek, S., and Boughanmi, A. (2020).
\newblock Impact of intellectual capital on the financing of startups with new
  business models.
\newblock {\em Journal of Knowledge Management}.

\bibitem[Peneder, 2008]{peneder2008problem}
Peneder, M. (2008).
\newblock The problem of private under-investment in innovation: A policy mind
  map.
\newblock {\em Technovation}, 28(8):518--530.

\bibitem[Ricciardi et~al., 2020]{ricciardi2020bayesian}
Ricciardi, F., Mattei, A., and Mealli, F. (2020).
\newblock Bayesian inference for sequential treatments under latent sequential
  ignorability.
\newblock {\em Journal of the American Statistical Association},
  115(531):1498--1517.

\bibitem[Rom{\'a}n et~al., 2013]{roman2013start}
Rom{\'a}n, C., Congregado, E., and Mill{\'a}n, J.~M. (2013).
\newblock Start-up incentives: Entrepreneurship policy or active labour market
  programme?
\newblock {\em Journal of Business Venturing}, 28(1):151--175.

\bibitem[Rosenbaum and Rubin, 1983]{rosenbaum1983}
Rosenbaum, P.~R. and Rubin, D.~B. (1983).
\newblock The central role of the propensity score in observational studies for
  causal effects.
\newblock {\em Biometrika}, 70(1):41--55.

\bibitem[Rubin, 1974]{rubin1974}
Rubin, D.~B. (1974).
\newblock Estimating causal effects of treatments in randomized and
  nonrandomized studies.
\newblock {\em Journal of educational Psychology}, 66(5):688.

\bibitem[Rubin, 1978]{rubin1978}
Rubin, D.~B. (1978).
\newblock Bayesian inference for causal effects: The role of randomization.
\newblock {\em The Annals of statistics}, pages 34--58.

\bibitem[Rubin, 1980]{rubin1980}
Rubin, D.~B. (1980).
\newblock Randomization analysis of experimental data: The fisher randomization
  test comment.
\newblock {\em Journal of the American statistical association},
  75(371):591--593.

\bibitem[Rubin et~al., 2006]{rubin2006}
Rubin, D.~B. et~al. (2006).
\newblock Causal inference through potential outcomes and principal
  stratification: application to studies with “censoring” due to death.
\newblock {\em Statistical Science}, 21(3):299--309.

\bibitem[Shane, 2009]{shane2009encouraging}
Shane, S. (2009).
\newblock Why encouraging more people to become entrepreneurs is bad public
  policy.
\newblock {\em Small business economics}, 33(2):141--149.

\bibitem[Wu and Carroll, 1988]{wu1988estimation}
Wu, M.~C. and Carroll, R.~J. (1988).
\newblock Estimation and comparison of changes in the presence of informative
  right censoring by modeling the censoring process.
\newblock {\em Biometrics}, pages 175--188.

\bibitem[Zhang and Rubin, 2003]{zhang2003estimation}
Zhang, J.~L. and Rubin, D.~B. (2003).
\newblock Estimation of causal effects via principal stratification when some
  outcomes are truncated by “death”.
\newblock {\em Journal of Educational and Behavioral Statistics},
  28(4):353--368.

\bibitem[Zhang et~al., 2008]{zhang2008evaluating}
Zhang, J.~L., Rubin, D.~B., and Mealli, F. (2008).
\newblock Evaluating the effects of job training programs on wages through
  principal stratification.
\newblock In {\em Modelling and Evaluating Treatment Effects in Econometrics}.
  Emerald Group Publishing Limited.

\end{thebibliography}
